# ON THE MASS OF A UNIVERSE WITH A QUANTUM STRING-LIKE BEGINNING

A. J. MEYER, II
*International Scientific Projects, Inc.*
*PO Box 3477*
*Westport Connecticut, 06880 USA*
*Email:* ajmeyer@InternationalScientificProjects.org

**Abstract**

In this paper the mass of a universe, which began as a gravitationally closed, maximally spinning, Planck density quantum string is derived. A universe beginning in such an initial state has been called a Super Spin Model universe.

The total mass, M, of the Super Spin Model universe is shown to be a function of only four fundamental parameters ℏ, c, G, e, such that: $M = \pi^2(\hbar c/G)^{1/2} exp(\hbar c/e^2)$.

The present paper primarily consists of a brief derivation of the above equation and a short synopsis of the Super Spin Model.

**Introduction**

This is a brief review of the Super Spin Model, wherein the relationship:

$$M = 4\pi^2(\hbar c/G)^{1/2} exp(\hbar c/e^2)$$

was established for a non-stationary axi-symmetric space-time, where $Mc^2$ is the space-time's total conserved energy.[1]

The present paper primarily consists of a brief derivation of the above equation and a short synopsis of the Super Spin Model (SSM).

At some point near the beginning, the universe apparently was in a super dense energy state of about $10^{92}$ gm·$c^2$/cm$^3$.

Other than introducing the somewhat ad-hoc mechanism of inflation, how could the nascent universe expand and overcome the intense gravitational fields that normally would reduce it to a state of permanent and complete gravitational collapse?

According to the SSM, the reason the universe expands is due to its initial rotating toroidal string−like structure − its topology. If the energy were distributed in the form of a maximally rotating Planck density string, then because of the initial balance of the centrifugal and gravitational forces, and due to its small surface area, the string would begin to evaporate into its "ergosphere" and in spite of its huge density would naturally begin to expand into its other, initially "compressed," dimensions. Such a string would already have been "pre-inflated" in one spatial dimension, while "compressed" in the others.

To reiterate, the SSM universe is postulated to have begun as a gravitationally closed, maximally rotating Planck density geon, i.e. a four-dimensional closed string with three of its four dimensions "curled up" or compacted within contiguous Planck areas of their locally mutually orthogonal surfaces. Such a geon,[2] with mass M and angular momentum magnitude J, has certain aspects that resemble the ring singularity of an extreme Kerr black hole. One may visualize this beginning state as a circle of light – a *geonic* pseudo ring singularity – a closed string of maximum energy density light consisting of Planck mass photons – "*primatons*." See [Wheeler (1955)] and [Meyer (1980)].

---

[1] Three additional independent equations for the mass 'M' of such non-stationary axi-symmetric space-times were also derived that led to linkages amongst: Newton's universal gravitational constant 'G', the fundamental quantum of electric charge 'e', the fundamental quantum of angular momentum – Planck's constant 'ℏ', the speed of light 'c' in vacuo and the electron and proton masses, $m_e$ and $m_p$. This set of connections, in turn, led to linkages among the four fundamental forces: the Gravitational, Electromagnetic, Strong, and Weak. These linkages are discussed in Meyer (1995).

[2] Geons are gravitational electromagnetic entities. They are regions consisting of electromagnetic radiation, which have sufficient energy density to become gravitationally confined. That is, light itself forms a sort of black hole. Wheeler developed this concept in 1954.

The SSM is partially inspired by the Kerr family [Kerr (1963)] of stationary solutions for the empty space, axi-symmetric, Einstein-Maxwell, source free equations. However, an expanded Bekenstein-Hawking thermodynamics is used which incorporates both the inner and outer event horizon areas as measures of entropy, thereby yielding complete consistency with the Third Law of Thermodynamics.[3]

By employing a continuous variation of parameters methodology along with a parsimonious set of initial boundary constraints, the stationary metric is extended to a non-stationary axi-symmetric space-time possessing non-stationary "evolving" event horizons.

Due to their parametric simplicity, the uncharged Kerr family provides a rich laboratory in which to perform "gedanken versuchen." This uncharged family is determined by just two parameters: E and **J**, where $E=Mc^2$ is the total energy of the space-time and **J** is its total angular momentum. This family is represented by the following metric:

$$ds^2 = (2R_g r - \sigma^2)du^2/\sigma^2 - 2aR_z^2 sin^2\theta d\phi du/\sigma^2 + R_\phi^4 sin^2\theta d\phi^2/\sigma^2$$
$$+ 2dudr - 2a sin^2\theta dr d\phi + \sigma^2 d\theta^2.$$

Where:
$R_g \equiv GM/c^2 \equiv$ Gravitational radius of space-time,
$a \equiv$ Space-time's specific angular momentum radius $= J/Mc$,
$\sigma^2 \equiv r^2 + a^2 cos^2\theta \equiv$ Space-time's rotation radial coordinate offset,
$R_z^2 \equiv 2R_g r$,
$\gamma^2 \equiv R_z^2 - r^2 - a^2$,
$R_\phi^4 \equiv (r^2 + a^2)^2 + \gamma^2 a^2 sin^2\theta$,
$Mc^2 =$ Conserved total energy of space-time,

And:
$J \equiv |\mathbf{J}| \equiv aMc \equiv$ Scalar value of space-time's angular momentum,

such that:
$-\infty \leq r \leq +\infty$,
$-\infty \leq u \leq +\infty$,
$0 \leq \phi \leq 2\pi$,
$0 \leq \theta \leq \pi$.

Note, there exist two non-negative event horizon radii, $r_+$ and $r_-$, which are the conjugate solutions of the equation: $\gamma^2(r) = 0$, which are: $r_\pm = R_g \pm [R_g^2 - a^2]^{1/2}$.
They can be re-written as:
$$r_\pm = R_g[1 + sin(\pm \Phi)] \Rightarrow$$
$$a^2 = r_+ r_- = R_g^2 cos^2(\Phi) \text{ and } r_+ + r_- = 2R_g.$$

Hence, the space-time's specific angular momentum radius, 'a' is parameterized in terms of M and $\pm\Phi$, so that:
$$a = a(\pm\Phi) = R_g cos(\Phi) = GM cos(\Phi)/c^2.$$

The angle '$\Phi$' is therefore a measure of the magnitude of the space-time's angular momentum (and/or expansion state). That is:
$$J(M, \pm \Phi) = a(\pm\Phi)Mc = R_g Mc\, cos(\Phi) = GM^2 cos(\Phi)/c = |\mathbf{J}(M, \pm \Phi)|.$$

As is easily seen, $\Phi = \pm 0$ implies a maximally rotating Kerr space-time and $\Phi = \pm \pi/2$ implies a static Schwarzschild space-time.

---

[3] It can be readily shown that the Third Law of Thermodynamics is satisfied if the area of the inner event horizon of a Kerr black hole is treated as a measure of the hole's negative entropy and, as usual, the outer event horizon area is treated as a measure of its positive entropy.

## The Description of the SSM Universe (U) as a Rotating *Non-Stationary* Axi-Symmetric Space-Time

One way to produce a model of a non-stationary axi-symmetric space-time, is to characterize the entire family of stationary uncharged Kerr solutions as the set of ordered pairs:

$$\{<M, \mathbf{J}>\} = \{<M, \mathbf{J}(\pm\Phi)> \ni [-\pi/2 \leq \Phi \leq \pi/2]\}.$$

Such a parameterization gives the entire family of Kerr solutions in terms of only two parameters 'M' and '$\Phi$' and only two universal constants 'G' and 'c.' That is any member of the Kerr metric family can be represented as the ordered pair: $<M, \mathbf{J}(\Phi)> = <M, GM^2 cos(\Phi)/c>$.

By treating $\pm\Phi$ as a dual-time dependent variable, it is possible to transform the Kerr metric for the *stationary family*, wherein each member contains a pair of *completed* inner and outer event horizons, into a *single dual-valued time-varying metric*. In other words, the *family* of metrics is transformed into a *single* metric function of $\Phi$, which governs the evolution of the *non-stationary incomplete* inner and outer event horizons, $H(r_\pm(\Phi),\theta_\pm(\Phi))$, that emanate from the "creation ring" and *grow* in both $\pm\Phi$ directions, starting from $\Phi = 0$, where: $r_\pm(\pm 0) = R_g$ and $\theta_\pm(\pm 0) = \pi/2$. The expansion trajectories of the event horizons' altitudinal "rims" are governed by: $\theta_\pm(\Phi) = \pi/2 \pm \Phi|$.

For the stationary metric, when $\gamma^2(r) = 0 \Rightarrow r_\pm = R_g \pm [R_g^2 - a^2]^{1/2}$, then $ds_\pm = 0$. However, this is not generally true when the metric coefficients are written as a function of the parameter $\Phi$. That is, $ds_\pm/d\Phi$ is generally $\neq 0$, for the hypersurfaces, $\gamma^2(r_\pm) = 0$, when $\pm\Phi$ is a dual-time dependent variable.

Rewriting the metric in terms of its past directed and future directed non-stationary event horizons,[4] we get:

$$ds_\pm^2 = \sigma_\pm^2 d\theta^2 - 2a sin^2\theta dr_\pm d\phi + 2dr_\pm du_\pm + sin^2\theta[R_{z\pm}^2 d\phi - a sin\theta du_\pm]^2/\sigma_\pm^2.$$

Where the boundary conditions of '**U**' are:

$$0 \leq r_- \leq R_g \leq r_+ \leq 2R_g,$$
$$0 \leq \theta_- \equiv \pi/2 - |\Phi| \leq \theta \leq \pi/2 + \Phi| \equiv \theta_+,$$
$$0 \leq \phi \leq 2\pi,$$
$$-\infty \leq u \leq +\infty.$$

And: $r_\pm = R_g[1 + sin(\pm\Phi)]$ are the *event horizon radii*,

$\sigma_\pm^2 = r_\pm^2 + a^2 cos^2\theta$ are the *rotation-offset radii*,

$R_{z\pm}(\Phi) = [2R_g r_\pm]^{1/2} = [r_\pm^2 + a^2]^{1/2} = R_g[2(1 \pm sin\Phi)]^{1/2}$ are the *radii of gyration* of the event horizons, and

$\Omega_\pm(\Phi) = c \, d\phi/du_\pm = -g_{u\phi}(\Phi)c/g_{\phi\phi}(\Phi) = ac/(R_{z\pm})^2 = c \, cos\Phi/[2R_g(1 \pm sin\Phi)]$ are the *dual event horizon rotation rates*.

For a constant M, one can see that the *moments of inertia* are: $I_\pm = M R_{z\pm}^2$,
since: $I_\pm \Omega_\pm = J(\pm\Phi) = a(\pm\Phi)Mc = R_g Mc \, cos(\Phi) = GM^2 cos(\Phi)/c = |\mathbf{J}(\pm\Phi)|$.

The above formulations are applicable to the subset **U** of the axi-symmetric space-time manifold,[5] **M**, such that **U** is isomorphic to the intrinsically non-stationary region, which is partially bounded by or swept out by the two "incomplete" event horizons with radii: $r_\pm = r(\Phi(t_\pm))$. Notice at $\Phi = \pm\pi/2$, the **U** becomes identical to a Schwarzschild black hole with complete and stationary event horizons. It is also critical to notice, that for constant $\Phi$ (the stationary case), the $r_\pm$ are the loci of null hyper-surfaces, but for variable $\Phi$ (the non-stationary case) the $r_\pm$ loci are null only where:

$$cos^2\theta = sin^2\Phi \Rightarrow \theta = \theta_\pm \equiv \pi/2 \mp \Phi.$$

The inner and outer event horizon areas are calculated by:

$$A(\pm\Phi) = \int_{(\pi/2-\Phi)}^{(\pi/2+\Phi)} \int_0^{2\pi} \sqrt{(g_{\phi\phi} g_{\theta\theta})} \, d\phi d\theta = 4\pi R_{z\pm}^2 sin(\pm\Phi) =$$

$$8\pi R_g^2 sin(\pm\Phi)[1 + sin(\pm\Phi)].$$

---

[4] See Meyer (1995)

[5] There might also exist an independent **anti M $\equiv$ - M,** simultaneously created with equal energy and angular momentum, but with opposite parity.

Therefore, the total *net* horizon area[6] for the SSM Universe is reckoned as:
$$A_U^{(2)}(\Phi) = A(+\Phi) + A(-\Phi) = 16\pi(R_g \sin\Phi)^2.$$
Moreover, the spatial volume[7] or hyper-surface area of the SSM U is calculated as:
$$A_U^{(3)}(\Phi) = 4\pi R_g^3 \sin\Phi(5\Phi - 3\sin\Phi\cos\Phi) \equiv V_U(\Phi).$$
We also find that the space-time 4-volume, (ignoring the √-1 coefficient) is:
$$A_U^{(4)}(\Phi) = (8\pi c/3) \, t_s(\Phi) R_g^3 \sin^2\Phi \, (3 + 2\sin^2\Phi - \sin^4\Phi).$$
Of course, the circumference $C_{\gamma\pm}$ is the one-dimensional "area":
$$A_U^{(1)}(\Phi) = 4\pi R_g.$$
A net time interval,[8] $t_s(\Phi)$, defined as the difference between the future directed, $u_+(\Phi)/c$, and past-directed, $u_-(\Phi)/c$, geodesic time displacements, is reckoned by:

$$ct_s(\Phi) \equiv [u_+(\Phi) - u_-(\Phi)] = R_g[-\ln(1-\sin|\Phi|) - \ln(1+\sin|\Phi|)] = -R_g \ln(1-\sin^2\Phi).$$

In general: $ct_s(\pm\Phi) = -R_g \ln(\cos^2\Phi)$. Note that at maximum expansion, when $\Phi = \pm\pi/2$, then:
$$ct_s(\pm\pi/2) = -R_g \ln[1-\sin^2(\pm\pi/2)] = +\infty.$$

Observe that the past-directed[9] time displacement: $t_-(\Phi) = R_g \ln(1+\sin|\Phi|)/c$, is always finite.

The expansion parameter '$\Phi$' grows in both directions away from an origin $\Phi \sim \pm 0$. That is, the space-time expands both ways in time.[10] In the beginning, the space-time or "creation ring" is a closed cosmic string with an *inflated* circumference. Since **J**, the angular momentum vector, is a cosine function of $\Phi$, then by increasing $|\Phi|$, the angular momentum magnitude is decreased and the space expands. As $\Phi$ approaches $\pm\pi/2$ the outer event horizon approaches the Schwarzschild radius and the inner event horizon radius approaches zero.

See **Figs. 1**, **2** and **3**, which are schematics that attempt to depict how the SSM universe evolves.

**The Initial Conditions**

Let $\Phi_\chi \approx \pm 0$ be the values of $\Phi$ when the density of the toroidal geonic string universe is equal to the Planck photon density.[11] This occurs when:
$$\rho \equiv M/V_U(\Phi_\chi) = M/4\pi R_g^3 \Phi_\chi^2 = \rho_\varpi = c^5/8\pi^2 G^2 \hbar$$
$$\Rightarrow \Phi_\chi = \pm\sqrt{\pi}/N_\varpi.$$

Therefore, for the small angle at creation $\Phi = \Phi_\chi$ one gets:
$$ct_s(\Phi_\chi) = -R_g \ln(1-\sin^2\Phi_\chi) \approx R_g \Phi_\chi^2 = \pi R_g/N_\varpi^2 = \pi\hbar/Mc \equiv ct_\chi,$$
which is *one half*[12] the Compton wavelength of the SSM universe. Note for a universe of mass $M \sim 10^{56}$ gm, the Compton time, or creation interval $t_\chi$ is about $10^{-104}$ sec.

---

[6] Note that at $\Phi = 0$, the net event horizon area is zero, which of course means that the Bekenstein-Hawking entropy is also zero.
[7] This spatial volume calculation is worked out in Meyer (1995).
[8] Meyer (1995), (2000)
[9] Whereas, the maximum past time is always finite, the maximum future time is always infinite.
[10] Notice that both directions and senses of the temporal dimension, i.e. time and anti-time began (were created) at $\Phi = 0$, along with the other spatial dimensions. It is simple but interesting to observe that anti-time trajectories cannot reach the "terra incognita" beyond or before the universe was created.
[11] In this paper, the Greek letter varpi '$\varpi$' denotes a Planckian parameter, so that the Planck mass is: $m_\varpi = (\hbar c/G)^{1/2}$, the Planck momentum magnitude is: $p_\varpi = m_\varpi c$, the Planck radius is: $r_\varpi = \hbar/p_\varpi$, the Planck wavelength is: $\lambda_\varpi = 2\pi r_\varpi$, the Planck time is: $t_\varpi = \lambda_\varpi/c$ and so forth.
[12] This might mean there were two universes created as a quantum fluctuation of the vacuum. That is, there might be another **U** created with opposite parity, so that the net angular momentum would be zero. In this paper, we shall just be concerned with one of them.

Nevertheless, immediately after the instant of creation at $t_\chi \sim 10^{-104}$ sec, the **U**'s *minimum* spatial dimension is already on the order of the Planck length. It therefore follows that the lowest bound of any future temporal increment must be on the order of the Planck time.

If the universe started as a non-rotating bubble membrane fluctuation, then because of the extreme "thinness" of the membrane, there is no apparent reason that such a non-rotating bubble would be stable. It should "flash back" into the vacuum from whence it sprang within its Compton time,[13] $t_c = 2\pi\hbar/Mc^2$.

However, in the SSM, the angular momentum of the maximally spinning string induces a coarser granularity by a factor of $N_\varpi$ upon both the spatial and temporal quanta of the fluctuation's active region. That is, the previously indivisible quanta become Planck sized.

Thus, the angular momentum of the creative process produces a real space-time, "inflating," by a factor of $N_\varpi$, the universe's original indivisible spatial and temporal quanta, i.e., its Compton length and time, to the Planck length and time, ($r_\varpi$ and $t_\varpi$), which become its new indivisible spatial and temporal quanta.

The rotating string topology thereby produces a "one-way valve" which "closes all the hatches" and stops the massive virtual fluctuation from "sinking" back into the vacuum within $\sim 10^{-104}$ sec. In other words, the Creator "writes a check" to the vacuum for about $10^{56}$ grams of energy and then "changes the *banking rules* before it can clear."

At creation, when $\Phi = \Phi_\chi = \pm\sqrt{\pi}/N_\varpi$:

$\exists$ a gravitationally closed, Planck density string of *length*[14] $\equiv C_{\gamma\pm} = 4\pi R_g$.

$\exists$ maximum electromagnetic energy density[15] $= \rho_\varpi c^2 = c^7/8\pi^2 G^2 \hbar$.

$\exists$ maximum action for the gravitationally closed space $= |\mathbf{J}(0)| = GM^2/c$.

$\exists$ maximum string tension $= T_\varpi = Mc^2/C_{\gamma\pm} = c^4/4\pi G$.

$\exists$ maximum vacuum polarization[16] energy $= \varepsilon_{evac\varpi} = e^2(c^3/\hbar G)^{1/2} = kT_{evac\varpi}$.

$\exists$ maximum photon energy $= \varepsilon_{\gamma\varpi} = (\hbar c^5/G)^{1/2} \equiv m_\varpi c^2$.

**Derivation of the String's Mass**

Along the string's ergosphere, the Unruh temperature [Unruh (1976)] is reckoned as:

$T_{evac\varpi} = \hbar\kappa_{evac\varpi}/2\pi ck$, where: $\kappa_{evac\varpi} = 2\pi\partial^2\mathbf{r}/\partial t^2|_\varpi = e^2 c/r_\varpi \hbar$ and $r_\varpi = \lambda_\varpi/2\pi$.

And when $|\Phi| \leq |\Phi_\chi|$, then: $r_\pm(\Phi) = R_g \pm \delta_r \approx R_g$ and $\theta_\pm(\Phi) = \pi/2 \pm \delta_\theta \approx \pi/2$.

In other words, a main postulate is that the initial fluctuation is a non-singular coherent toroidal electromagnetic disturbance of Planck density, $\rho_\varpi$. This fluctuation is formally identical to a toroidal geon, or a closed circular string of mass $N_\varpi$ in Planck units, with total initial angular momentum magnitude:

$$J(0) = [\mathbf{J}(0)\cdot\mathbf{J}(0)]^{1/2} = \hbar[N_\varpi^4 + 2N_\varpi^2 + 3]^{1/2} \approx \hbar N_\varpi^2 = R_g Mc,$$
where: $N_\varpi \equiv M/m_\varpi$.

After a little algebra, one finds that the initial volume of the six dimensional phase space occupied by the string-like universe at $\Phi_\chi$ is:

$$V_0^{(6)}(\Phi_\chi) = \int dV^{(3)} \int dp^{(3)} = 8\pi^2 R_g r_\varpi^2 p_\varpi^3 = 8\pi^2 N_\varpi \hbar^3.$$

However, the total phase space volume at $\pm\Phi_\chi$ is the sum of the volumes of all cubic action compartments $\hbar^3$. This volume is calculated as:

---

[13] See discussion in Meyer (1995)

[14] $\forall \Phi \ni \gamma(r_\pm(\Phi)) = 0$, $C_{\gamma\pm} = \int_0^{2\pi}[g_{\phi\phi(\gamma=0)}]^{1/2}d\phi = 4\pi R_g$ = circumference of *both* inner and outer event horizons at the "equator" (the space orthogonal to the axis of symmetry, $\mathbf{J}(0)$, i.e. $\theta = \pi/2$) Note, *the circumference is constant for all $\Phi$ and hence, is independent* of the event horizon radii.

[15] See the discussions about maximum energy density electromagnetic quanta in Meyer (1980), (1995).

[16] See Meyer (1995).

$$V_S^{(6)}(\Phi_\chi) = [(\mathbf{J}_x(\Phi_\chi) \pm \hbar) \times (\mathbf{J}_y(\Phi_\chi) \pm \hbar)] \cdot (\mathbf{J}_z(\Phi_\chi) \pm \hbar).$$

Now, since the metric is axi-symmetric, we can choose: $\mathbf{J}_x(\Phi_\chi) = \mathbf{J}_y(\Phi_\chi) = \mathbf{0}$ and $|\mathbf{J}_z(\Phi_\chi)| = R_g Mc$. It then follows that: $V_S^{(6)}(\Phi_\chi) = (R_g Mc \pm \hbar)\hbar^2 = N_\varpi^2 \hbar^3 \pm O(\hbar^3)$. Moreover, the ratio of the string's initial sub-phase space volume, $V_0^{(6)}(\Phi_\chi)$, (*with all photons in the Planck state*) to the total phase space volume $V_S^{(6)}(\Phi_\chi)$, should be equal to the Bose-Einstein expectation number, $\langle n_\varpi \rangle$.

That is: $\langle n_\varpi \rangle = V_0^{(6)}(\Phi_\chi)/V_S^{(6)}(\Phi_\chi) = 8\pi^2 N_\varpi \hbar^3/N_\varpi^2 \hbar^3 = 8\pi^2/N_\varpi =$
$$2/[exp(\varepsilon_{\gamma\varpi}/kT_{evac\varpi}) - 1] = 2/[exp(\varepsilon_{\gamma\varpi}/kT_{evac\varpi}) - 1] = 2/[exp(\hbar c/e^2) - 1)].$$

Since, $\hbar c/e^2 = \alpha^{-1} \approx 137$, we obviously can neglect the '-1' term and therefore get:
$$N_\varpi = 4\pi^2 exp(\hbar c/e^2) = 4\pi^2 exp(1/\alpha) = M/m_\varpi.$$

Therefore, the mass, M, of this rotating string universe is reckoned as:
$$M = 4\pi^2 (\hbar c/G)^{1/2} exp(\hbar c/e^2) = 2.8062 \times 10^{56} \text{ gm}.$$

**Entropy Formulation**

It has been found that the thermodynamics of stationary black holes becomes consistent with classical thermodynamics by introducing an extended interpretation of the "Bekenstein-Hawking"[17, 18] entropy formula, by augmenting it to include the inner event horizon area as a measure of "negentropy."[19]

By also extending this augmentation to the incomplete and evolving event horizons of the non-stationary SSM **U**, the net entropy of **U** as a function of $\Phi$ then becomes:
$$S(\pm\Phi) = 4\pi k (N_\varpi \sin\Phi)^2 \geq 0, \forall \Phi.$$

It is easy to see that entropy will increase along both positive and negative temporal directions as both temporal displacements increase away from the origin in both positive and negative time.

There is a low initial entropy value upon the creation of the Planck density string[20] at $\Phi = \Phi_\chi$. This value is:
$$S(\Phi_\chi) = 4\pi k(N_\varpi \sin\Phi_\chi)^2 = 4\pi^2 k,$$
where $\Phi_\chi = \pm\sqrt{\pi/N_\varpi}$ and 'k' is Boltzman's constant.

Since $ct_s(\pm\Phi) = -R_g ln(1 - \sin^2\Phi_\chi)$, then as $\Phi \to \pm\pi/2$ and $t_s \to \infty$, the **U** approaches its Schwarzschild limit, and the net event horizon area and net entropy approach their maxima.[21] From the above equation for $S(\pm\Phi)$ the maximum entropy is therefore:
$$S(\pm\pi/2) = 4\pi k N_\varpi^2 = 64\pi^5 k\, exp(2/\alpha) = 1.51287 \times 10^{139} \text{ (erg/°K)}.$$

---

[17] Bekenstein (1973)
[18] Hawking (1974)
[19] Since the dual gravitational "*strengths*" or "*accelerations,*" $\kappa_\pm$, for stationary Kerr black holes are reckoned by: $\kappa_\pm = \pm c^2(R_g^2 - (J/Mc)^2)^{1/2}/2R_g[R_g \pm (R_g^2 - (J/Mc)^2)^{1/2}] = 0$. When $R_g = J/Mc$, both event horizon temperatures are calculated as: $T_\pm = \hbar\kappa_\pm/2\pi kc = 0$. However, the un-augmented Bekenstein-Hawking entropy is: $S_{BH} = 2\pi kGM^2/\hbar c \gg 0$, which leads to a violation of the Third Law.
[20] In the SSM, since the space-time is non-stationary and its string-like initial state is far from equilibrium, the initial net entropy is near zero, whereas its initial temperature is very large.
[21] The excess angular momentum energy, $\Delta E_J = Mc^2[1-1/\sqrt{2}]$, that is not taken up by the expansion, see Christodoulu (1970), is conjectured to be taken up by the generation of "flywheels" surmised to be a species of neutrinos, see Meyer (1995). Some of the excess may also be stored in the rotational energy of galaxies, stars and other spinning and rotating objects.

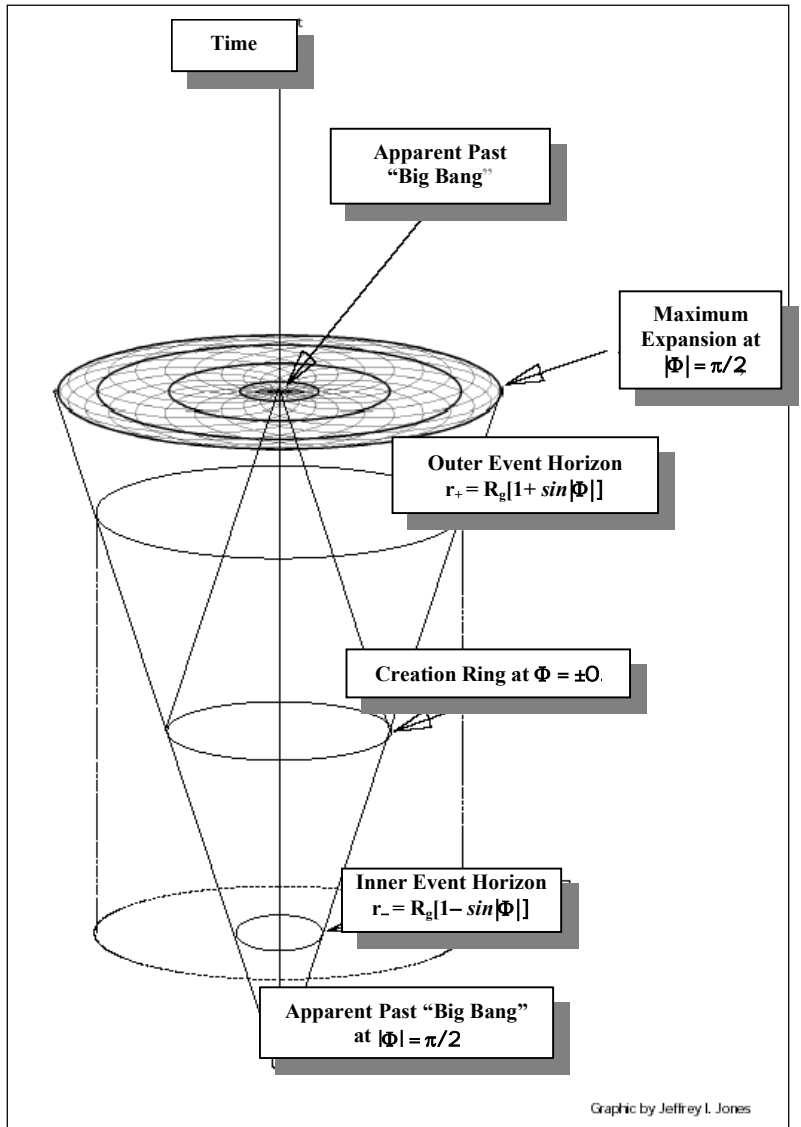

**Figure 1. Portrait of evolution of SSM universe**

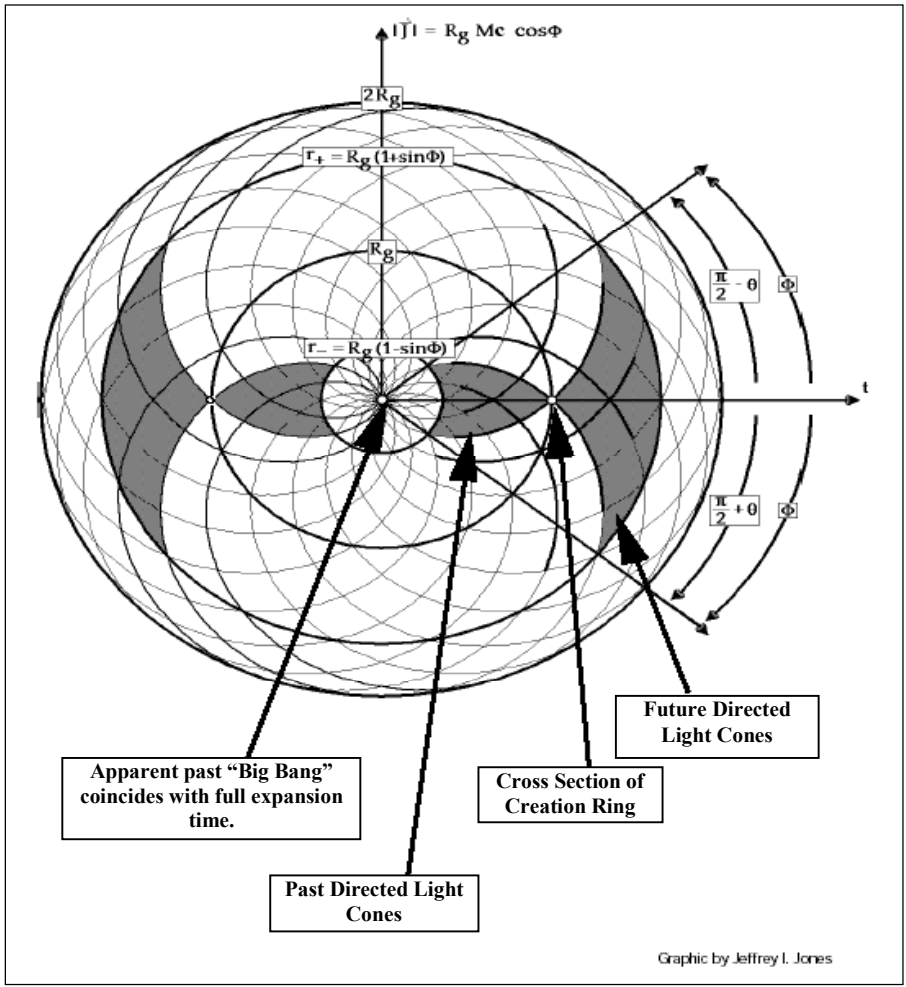

**Figure 2.** Profile of future directed and past directed light cones starting at creation ring as projected onto the following space: {< ±t, θ, φ > ∋ (π/2– Φ ≤ θ ≤ π/2 +Φ)∧(φ = constant )}.

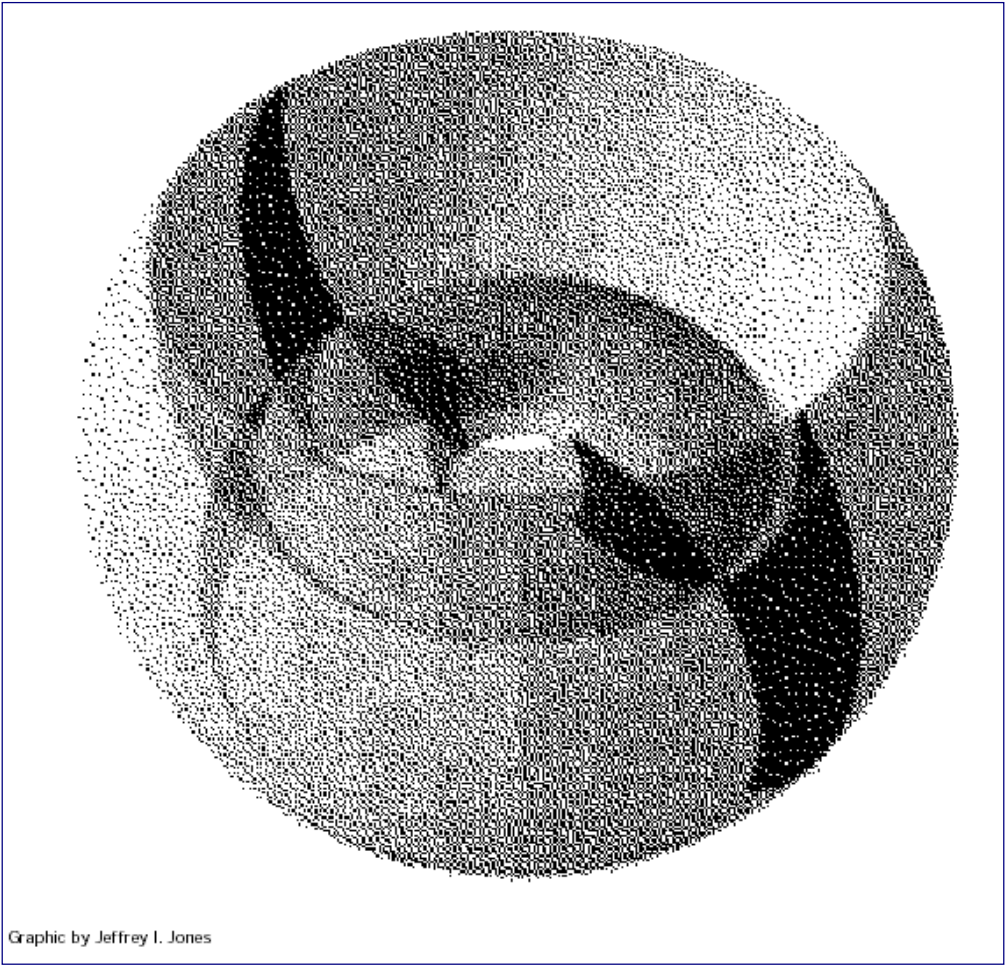

**Figure 3   (Solid View of Figure's 1 & 2)**

This illustrates the loci of future and past-directed light cones, starting at creation ring as projected in the following space:{< ±t, θ , φ > ∋ ( π/2– Φ ≤ θ ≤ π/2 +Φ ) ∧ (0≤ φ ≤ 2π)}.  The darkest region is a two-dimensional {< ±t, θ >} profile of light cones as shown in Figure 2.  Only two-dimensional projections of three-space are shown in the above solid.

**ACKNOWLEDGEMENTS**

I am very grateful for the friendship, inspiration, and thoughtful encouragement that Professor Behram Kursunoglu gave me over many years. I miss him greatly. I also wish to thank Professor Arnold Perlmutter for much helpful advice, Professor Jeffrey I. Jones, of Queensland Technological University for graphics and technical expertise, and to my wife Connie, and daughters Michelle and Eva, for their prayers and support in getting this paper and presentation together. Above all, I wish to acknowledge and thank the eternal *"God (Who) used beautiful mathematics in Creating the World!"* as P.A.M. Dirac proclaimed.